# Monet: A User-oriented Behavior-based Malware Variants Detection System for Android

Mingshen Sun, Xiaolei Li, John C.S. Lui, *Fellow, IEEE, ACM*, Richard T.B. Ma, Zhenkai Liang

*Abstract*—Android, the most popular mobile OS, has around $78\%$ of the mobile market share. Due to its popularity, it attracts many malware attacks. In fact, people have discovered around one million new malware samples per quarter [1], and it was reported [2] that over $98\%$ of these new malware samples are in fact "*derivatives*" (or variants) from existing malware families. In this paper, we first show that runtime behaviors of malware's core functionalities are in fact similar within a malware family. Hence, we propose a framework to combine "*runtime behavior*" with "*static structures*" to detect malware variants. We present the design and implementation of MONET, which has a client and a backend server module. The client module is a lightweight, in-device app for behavior monitoring and signature generation, and we realize this using two novel interception techniques. The backend server is responsible for large scale malware detection. We collect 3723 malware samples and top 500 benign apps to carry out extensive experiments of detecting malware variants and defending against malware transformation. Our experiments show that MONET can achieve around $99\%$ accuracy in detecting malware variants. Furthermore, it can defend against 10 different obfuscation and transformation techniques, while only incurs around $7\%$ performance overhead and about $3\%$ battery overhead. More importantly, MONET will automatically alert users with intrusion details so to prevent further malicious behaviors.

## I. INTRODUCTION

ANDROID is a mobile operating system from Google and it powered mobile devices dominate around $78.7\%$ of the smartphone OS market in the first quarter of 2016 [3]. Android applications (apps for short) can be downloaded not only from the Google's official market Google Play, but also from third-party markets [4], [5], forums [6] and web sites. Although Google Play scans any uploaded apps to reduce malware [7], other markets/sites usually do not have sufficient malware screening, and they become main hotbeds for spreading Android malware. As a result, Android attracts millions of malware. It is reported that $97\%$ of mobile malware is on the Android platform [8].

Android provides various security mechanisms, such as the permission mechanism [9] and app verification [10]. The permission mechanism constrains functionalities of an app. Apps can only use permissions which are explicitly declared in their manifest files. When installing an app, users can review the requested permissions to decide whether to install the app or not. The permission system makes it difficult for attackers to obtain arbitrary privilege, but it does not help if the user accepts dangerous permissions requested by malware (and unfortunately, many users do exactly that). In addition, because of the permission abuse problem [11]–[13], malware can still find its way to attack many Android devices. Furthermore, researchers also propose a number of novel attack methods [14]–[18] targeting Android.

Malware detection is the key to provide Android security. Due to the difference in architectures, application structures and distribution channel, Android is very different from traditional platforms, hence conventional detection methods cannot be easily adapted to Android systems. To detect Android malware, a number of systems were proposed by industries and research communities. A widely deployed solution is to scan apps in the Android application market, i.e., the Bouncer scanner [7] in the Google Play Store. This helps to reduce (but not eliminate) malware in the Google Play market. However, due to the openness of the Android ecosystem, users often install apps from other markets or directly download from other sites (e.g., web forums). Hence, it is important to have in-device detection systems to target malicious apps.

Broadly speaking, there are two types of in-device malware detection systems. The first one is to perform static malware detection. This type of systems [11], [19]–[21] uses static information such as API calling information and control flow graphs to generate signatures for detection. For example, anti-virus engines will scan files in apps after their installation. However, studies [22], [23] have shown that these types of anti-virus engines can be easily bypassed using transformation attacks (i.e., code obfuscation techniques like package name substitution and reflection technique). Furthermore, sophisticated signature generation and signature matching techniques based on control flow analysis incur considerable computation overhead, and consume energy on mobile devices which have limited battery resource, preventing them from being adopted as in-device detection systems.

The second type of in-device detection system is the dynamic intrusion prevention system, as seen in several products [24]–[26] and research studies [27]–[29]. These systems work in the background and monitor apps at runtime. Once they discover any suspicious behavior, a notification will popup to alert the users. Note that suspicious behaviors are usually based on sensitive APIs. Many benign apps (e.g., text message management apps) may also invoke these APIs (e.g., sending text message API) for legitimate reasons. Therefore, this type of systems may introduce false alerts and makes intrusion notifications annoying and less preferable. Moreover,

Mingshen Sun and John C.S. Lui are with the Department of Computer Science and Engineering, The Chinese University of Hong Kong. Part of this work was done during Mingshen's internship at National University of Singapore. Email: {mssun, cslui}@cse.cuhk.edu.hk.

Xiaolei Li, Richard T.B. Ma, and Zhenkai Liang are with the School of Computing, National University of Singapore. Email: wenjie011369@gmail.com, {tbma, liangzk}@comp.nus.edu.sg.

a study [27] also shows that existing products in the market can be easily circumvented.

According to a survey [2], it was reported that over $98\%$ of new malware samples are in fact derivatives (or variants) from existing malware families. These malware variants use more sophisticated techniques like dynamic code loading, manifest cheating, string and call graph obfuscation to hide themselves from existing detection systems. Although these techniques can help malware to hide their malicious logic, we observe that the "*runtime behaviors*" of malware's core functionalities, such as unauthorized subscription of premium services or privilege escalation at runtime, remain unchanged. The runtime behaviors of a new malware variant and its earlier generation are usually very similar. A detection system based on *runtime behaviors* of malware will be able to detect most malware and their variants more reliably. In addition, the static structures of the malware are often similar within a malware family.

With this observation, we present the design and implementation of MONET, an Android malware detection system that combines "*static logic structures*" and "*dynamic runtime information*". MONET consists of a client module and a backend server module. The client module is a lightweight, in-device app for malware behavior monitoring and signature generation using two novel interception techniques, while the backend server module is responsible for malware signature detection. Our system can accurately describe the behaviors of an app to detect and classify malware variants and defend against obfuscation attacks. We focus on classifying malware based on their behavior similarity. The MONET's client module can be easily deployed on any Android mobile device. Moreover, it has low computational overhead and low demand on battery resources. Specifically, we make the following contributions:

- We design and implement a runtime behavior signature which can represent both the logic structures and the runtime behaviors of an app. Our runtime behavior signature is effective to detect malware variants and transformed malware.
- We implement a lightweight, in-device malware detection system, for Android devices. We propose two novel interception techniques, and show that it is easy to deploy and it provides informative alerts to users.
- We implement the solution, and demonstrate its effectiveness and its low overhead, both on CPU and battery resources.

The rest of the paper is organized as follows. Section II introduces the necessary background on Android. In Section III, we present the design of runtime behavior signature. In Section IV, we describe the MONET system, and the implementation details. In Section V, we evaluate the effectiveness and performance of MONET. Section VI presents the related work and the conclusion is given in Section VII.

## II. BACKGROUND

In this section, we introduce the essential background knowledge of Android malware variants and evaluation. We also discuss the intent interface and binder mechanism, which are important knowledge needed to design our interception techniques.

### A. Malware Variants and Evolution

To circumvent detection and to quickly deploy malware, hackers usually *do not* develop new malware from scratch, but rather improve existing logic or add new malicious logic into existing malware. They also repackage malware using disassembled tools [30], [31] to disassemble a benign app, and inject it with malicious logic, then repackage it as a new but malicious app. We call a set of malware with similar logic as a *malware family*. Moreover, if anti-virus engines can successfully detect these malware, malware writers will update parts of the logic of the original malware using some obfuscation techniques. These newly generated malware will have similar behavior as the original one. We call these malware as a "*variant*" within this malware family. According to a report [2], many Android malware samples are variations of existing malware. For example, the `DroidKungfu` family has four variants. They use native code, string obfuscation and encryption to make the malware more complicated and difficult for detection. Studies [22], [23] have shown that using simple transformations, anti-virus engines can be bypassed easily. We call the static and automatic transformation techniques such as string obfuscation, inserting junk instructions, renaming class names, as "*transformation attacks*". Therefore, detecting malware variants and defending against transformation attacks are challenging problems.

### B. Intent & Binder Mechanism

There are four types of components in an Android app. They are *activity*, *service*, *content provider* and *broadcast receiver*. An activity represents a screen on the devices which can interact with users. A service is a long-running background component which does not have a user interface, and their functions are to support tasks running in the background (such as playing music). Android provides many system-level services in the framework layer, for example, the activity manager and the SMS manager. Developers can also define services in their apps to provide functions for other apps. Content providers manage structured data such as SQLite database for apps. Broadcast receivers listen to events from other components such as boot completed events and SMS received event.

Because each component has individual functionalities and is isolated from other components, Android provides an interface which is called *intent* to connect these components. An intent is a messaging object which facilitates a component to request action from another component. Normally, one component can use intents to start an activity, start a service or deliver a broadcast. There are two types of intents: *explicit intent* and *implicit intent*. Explicit intent can start a component by specifying a full class name. For instance, knowing the names of classes, developers can use an explicit intent to start an activity or service in their own apps. Instead of explicitly declaring the name, implicit intent does not need the name of a component. Implicit intent can declare a general action to perform. Other components which are capable of performing such actions will handle this intent. For example, if an app wants to make a phone call, it can use an implicit intent with a dialing action (i.e., `android.intent.action.DIAL`) to start a dialer activity. However, if there is more than one dialer app, the system will popup a dialog for users to choose.

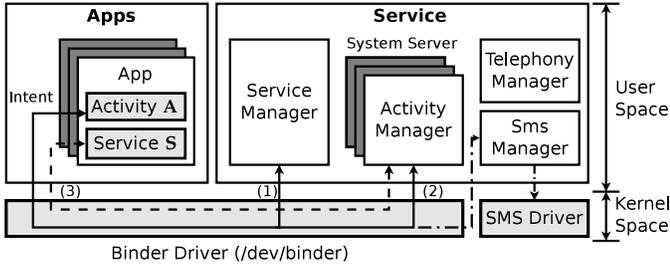

Fig. 1. Intent and binder mechanism.

From the operating system's perspective, one intent call involves three steps, which we illustrate in Figure 1. For instance, activity $A$ in an app wants to start the service $S$ using intent. Firstly, $A$ will request *Service Manager* to provide the address of the *Activity Manager* which is responsible for the activity related operations (e.g., starting activities and services). Then, $A$ will request Activity Manager to start the service $S$. In the final step, Activity Manager will tell this app to start the service $S$.

Because each app runs in a sandbox within an Android system, components belonging to different apps cannot directly communicate with each other in user space. But instead, Android system provides a kernel driver which is called the *binder* in kernel space for inter-process communication. We want to emphasize that intent is a high level abstraction in the application framework layer, and the implementation of intent utilizes binder driver in the kernel layer. Figure 1 illustrates the work flow of the intent call in the previous example. All the communications in the above mentioned three steps need to go through the binder driver. We call a binder communication as a *binder transaction*. There are several attributes in each binder transaction. *Binder descriptor* is a string which represents the target of this transaction. *Transaction code* is an integer indicating the action of this transaction. For instance, in the binder transaction from an app to the Activity Manager for starting an activity, the descriptor is `android.app.IActivityManager` and transaction code is 3. Besides the intent call, other APIs which need inter-process communications also utilize the binder mechanism. For example, to send a text message, an app will use the binder to request the SMS Manager to send a message through the SMS driver. In summary, binder calls can represent *all inter-process communication* including the intent calls between apps.

### III. SYSTEM DESIGN

In this section, we first state our problem, and then we discuss the system design of MONET, in particular, the design on the runtime behavior signature generation and the malware detection algorithm.

#### A. Problem Statement

One way to quickly mutate an Android malware is to use obfuscation methods to transform original codes to hide its malicious logic. Conventional methods for PC cannot be directly adapted to Android. Existing in-device solutions have limited capability to recognize malware, especially under the constraint of CPU resources and battery power. Our aim is to design a new and novel user-oriented approach for malware detection to achieve the following goals: (1) *resistant to malware variants and transformation attacks*, (2) *user-oriented and easy to deploy*, and (3) *highly efficient and scalable to detect large number of malware variants*.

- **Resistant to Malware Variants and Transformation Attacks.** MONET should detect malware variants which have similar runtime behavior. In addition, the transformation of static features such as package name, string and instruction order should not affect our detection results.
- **User-oriented and Easy Deployment.** MONET's client module is designed for common mobile device users rather than app marketplace to prevent malware. It should be easy to deploy, e.g., without modifying existing Android firmware. Moreover, after installing MONET on a mobile device, it should not consume much battery resource.
- **High Efficiency and Scalability.** After generating the signature, MONET's client module will send the information to the MONET's backend server for signature detection. The backend server needs to be efficient and scalable to support a large number of real time signature detection requests.

We like to mention that many current user-oriented anti-virus software programs only rely on static signatures which are generated from disassembled codes and other static resources (e.g., package names or unique strings within a malware family). In addition, many current dynamic analysis systems are designed only for assisting researchers to better understand the dynamic behaviors of malware. The current in-device intrusion prevention systems cannot determine the maliciousness of suspicious apps for users. Furthermore, mobile devices usually have constrained battery and computation resources, so conventional host-based intrusion prevention systems may not be appropriate.

#### B. Overview of Monet

Our system, MONET, determines the runtime behavior signature of malware, and it combines both the static logic structure and the runtime information. Runtime behavior is difficult to change, and this provides additional information for us to perform effective malware variant detection. Using this runtime behavior signature, MONET can detect malware variants and defend against malware transformation attacks. We design two interception techniques to realize our system so that users can easily install the MONET's client module on Android devices to provide malware protection.

MONET uses the following four steps to extract runtime information to perform malware detection: (1) *static behavior graph generation*, (2) *runtime information collection*, (3) *runtime behavior signature generation*, and (4) *signature detection*. Figure 2 illustrates the work flow of MONET to detect malware on Android devices.

When users install a new app on their devices, MONET monitors the installation event in the background, and extracts the static information including component information from the app's manifest file and static logic from the disassembled codes. Then, MONET generates a static behavior graph based on the static structure of the app before launching the app. After launching the app, MONET monitors and collects runtime information including binder transactions as well as some important system calls (e.g., `socket()` and `execve()` system call). If the system detects an intrusive action, it will popup a warning dialog to alert the user about the suspicious actions. If the user cannot determine the maliciousness of this action, the system will conduct further malware detection. MONET generates a runtime behavior graph for this app using the static behavior graph and the collected binder call information, and suspicious system calls will also be recorded for detection. Finally, MONET uses both the runtime behavior graph and the suspicious system call set as the *runtime behavior signature*, which is sent to the backend detection server for further analysis. The MONET's backend detection server, it will match any uploaded signature with existing malware signatures in the database, and return the result to the mobile device and notify users about the detection result.

### C. Runtime Behavior Signature

MONET uses runtime behavior signature (RBS) for malware detection. Runtime behavior signature includes both the *runtime behavior graph* (RBG) and the *suspicious system call set* (SSS). RBG contains not only the high level logic structure of an app, but also describes the calling actions among these logic structures at runtime. SSS contains execution information of sensitive system calls at runtime.

RBG is one of the basic elements for our behavior-based detection system. An RBG of an app is a directed graph over a set of app components and system components with two sets $\mathcal{C}$ and $\mathcal{B}$. $\mathcal{C}$ represents a set of app components which are all components used within an app and system components which are system services used, and $\mathcal{B}$ represents a set of binder calls. The set of vertices corresponds to the components in $\mathcal{C}$. The set of edges corresponds to the binder calls between two vertices in $\mathcal{B}$. The label of vertex contains the corresponding components names and properties. The label of edge consists of binder transaction code representing the calling purpose and the binder content containing essential information. For the implicit intent call in the RBG, because we do not know which component will handle the action of this intent, we treat this action as a node in the RBG. The property in the vertex label of a component indicates whether a node is an app component or a system component. In summary, because RBG describes the high-level logic structure within an app and the runtime interactions with other functional system components, we can use an RBG for behavior-based malware detection.

To further explain runtime behavior graph, we use an RBG of a malware (`o5android`) as an example to illustrate the details of RBG. This malware will register itself as a device administrator to prevent uninstallation, and it also uses the Google Cloud Messaging services to communicate with its command-and-control server to avoid detection. Figure 3 illustrates a part of the RBG of this malware. The black circles in the graph represent app components (i.e., the properties of the nodes) in the malware, and beside the nodes are the names of the nodes (i.e., the class names of the components). The white circles, on the other hand, represent system services which were requested by the malware at runtime, and the names of nodes are descriptors representing the system services. A link between two nodes implies a binder call between two nodes. The label of the link contains the transaction code and content of a binder call. In the left oval of the graph, there is a binder call from `com.google.elements.AdminActivity` to `android.app.action.ADD_DEVICE_ADMIN`. The code 3 represents an action to start an activity. Because the malware uses implicit intent to start the device administration app, the intent action is treated as a vertex in the RBG, which is the white node in the left oval. This part of the RBG describes a malicious behavior of the malware, which is registering the service as a device administrator. In the right dotted oval, there are two nodes and a link calling from `com.google.elements.MainActivity` to `com.google.android.c2dm.intent_REGISTER`. The behavior represented in this dotted oval is to initiate the Google Cloud Messaging service. We will illustrate the generation method of RBG in the following subsections.

RBG utilizes the specific app structure and communication mechanism for Android to record runtime behaviors. RBG contains two pieces of important information. The first one is the calling relation between components inside an app or what we call the logic structure, e.g., Activity `MainActivity` starts the service `AdminService`. The second component is what we call the runtime behaviors, e.g., Activity `MainActivity` obtains the device's unique ID through a telephony manager. Combining the logic structure and the runtime behaviors, RBG can accurately describe the characteristics of a malware. This is fundamentally different from existing static approaches [32] which simply use static features for malware detection. Next, we further elaborate how to use an RBG as a malware signature for detection.

**Role of Suspicious System Call Set (SSS):** SSS is a set of potentially dangerous system calls. For example, the system call includes `socket` and `execve` because malware can use `socket` to download malicious executable files and use `execve` to launch those programs. Firstly, malware may use socket (i.e., network) to communicate with the command and control server. MONET will capture the address of the connected server. Secondly, some trojans will execute root tools at runtime to gain root access and privilege. For example, `DroidKungfu` is a trojan malware which will execute the `secbino` binary to exploit system vulnerabilities. Because we can only obtain the calling process (i.e., app) rather than calling component of system calls in the kernel layer, we separate SSS as another element of runtime behavior signature and record them in SSS at runtime.

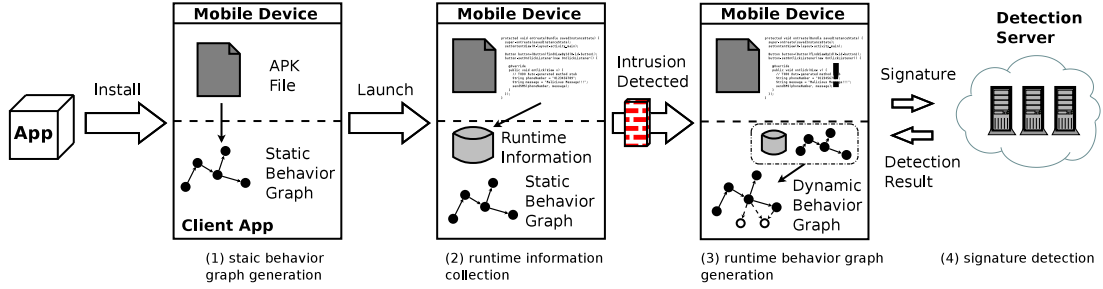

Fig. 2. Overview of MONET. Runtime behavior signature will be generated through static behavior graph generation, runtime information collection and runtime behavior graph generation.

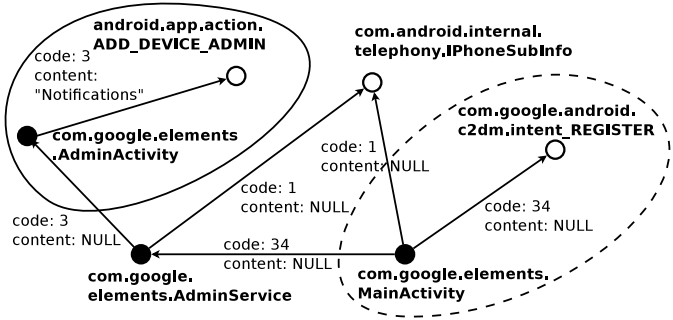

Fig. 3. Example of runtime behavior graph.

Together, both RBG and SSS constitute the runtime behavior signature of the app and we use it for malware detection. There are several reasons that RBG and SSS are suited as a basis for malware detection. Firstly, every component in an app has to use the binder mechanism to communicate with other components. So binder calls can accurately represent apps' runtime behavior. Also, for network behavior and binary execution, SSS can capture these suspicious system calls as supplementary runtime behaviors. Secondly, logic structures of a malware family are usually very similar. Although malware may use static obfuscation methods to avoid detection by static analysis, malware variants have similar run-time behavior with the original malware. Therefore, with an accurate representation of static structure and runtime behaviors, RBG and SSS can be used as a runtime behavior signature to detect malware variants and transformation attacks.

To generate an app's RBG, we need to extract the logic structure and the runtime behaviors. One can extract the app component information from the disassembled resources. However, we also need to execute the app to obtain the calling relation between components. Moreover, the calling relations rely on the execution routines of an app. To accomplish this, we propose to first use the static behavior graph (SBG), which can represent the static logic structure before launching the apps. In essence, SBG is a simplified RBG which only includes the app components and their static calling relation. In summary, SBG describes the *skeleton* (i.e., logic structure) of an app, and *connections* within the skeleton are provided by the runtime information, which we obtain from RBG. Specifically, there are two phases to generate an app's RBG. They are: (1) *static behavior graph generation* and (2) *runtime behavior completion*, which we explain as follows.

**(1) Static Behavior Graph Generation:** To generate an RBG, we first use the static information to generate the static behavior graph (SBG). SBG is a subgraph of RBG, but it does not contain runtime information. There are two steps to generate SBG. The first step is to extract app components from the app's manifest file (i.e., `AndroidManifest.xml` file). The second step is to find intent calls between components, i.e., one app component which starts another app component.

Note that for the second step, due to the limitation of static data-flow analysis, it is impossible to find all intent calls. For example, a malware can hide an intent call within a native code or obfuscate action string in the implicit intent call. Moreover, traditional static analysis methods impose high computation complexity. MONET uses an alternative method to statically extract all intent calls. Firstly, MONET will use the disassembled code to generate the control flow graph (CFG) for each class. Secondly, it searches all intent call methods (i.e., `startActivity` and `startService` APIs) in the CFG. Because there are several attributes in these intent call methods to indicate the caller and target, we can then keep track of these variables. Here, we use the reaching definition algorithm [33] to locate the caller and target. Lastly, we can determine an intent call and add a link in the SBG.

We want to point out that a full CFG and reaching definition analysis for an app will cost a lot computation resource. This is not feasible for battery constrained mobile devices. Therefore, we build a CFG and use the reaching definition algorithm only within a component class. For other binder calls which cannot be found by the SBG generation process, we can obtain them at runtime.

Figure 4 depicts an example of statically finding an intent call, which initiates from the activity $A$ to the activity $B$. We first locate the `startActivity` API call. The parameter `i` is the intent object. Then, by using the reaching definition algorithm, we can find the definition of `i`. Note that `i` is defined by the intent constructor. The parameters of the intent constructor are the caller class and the target class of an intent call. Therefore, we locate the caller variable (i.e., `v1`) and target variable (i.e., `v2`) of this intent call from the constructor method of intent. Then, we find the definitions of `v1` and `v2`.

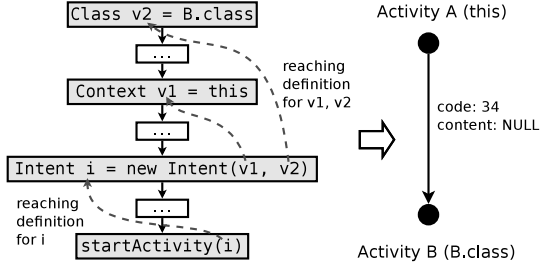

Fig. 4. Data-flow analysis for generating the static behavior graph (SBG).

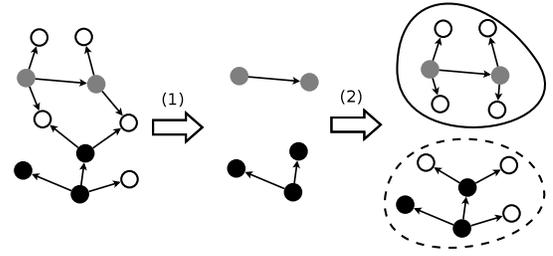

Fig. 5. The graph decoupling process.

Lastly, the system determines an intent call from the activity $A$ (i.e., this) to activity $B$, and this edge will be added into the SBG of this app. Using the above algorithm, most of the intent calls can be found and added to the SBG, which represents the skeleton of the app. Because we only perform reaching definition algorithm within each component logic, if definitions reside in other classes, we cannot locate this binder call. Moreover, some binder calls may be hidden inside native code. Therefore, the remaining calls will be recorded at runtime.

**(2) Runtime Behavior Completion:** Because SBG is based on static resources, it only possesses limited logic structure information. For example, malware samples may hide malicious logic by obfuscation and reflection techniques. To gain these hidden logic, we should capture runtime information. After executing the app at runtime, MONET can collect runtime binder calls. Then MONET will use these calls to complete the SBG and generate an RBG. After generating the RBG, which is a part of the signature of the suspicious app. MONET will send it and SSS to the backend detection server for malware detection. In Section IV, we will discuss in detail how we implement the runtime behavior collection process in MONET.

### D. Malware Detection

When the MONET's backend detection server receives the uploaded runtime behavior signature of a suspicious app, it will execute the signature matching algorithm to determine if this suspicious app is a malware. The detection algorithm involves three parts: (1) *graph decoupling*, (2) *malware signature generation* and (3) *signature matching*.

**(1) Graph Decoupling:** Because repackaged malware contains both benign logic and malicious logic, we need to perform a *graph decoupling* for all uploaded RBG to separate this logic for malware detection. Figure 5 illustrates the process of graph decoupling. Suppose we have an RBG of a repackaged malware. There are two steps to achieve graph decoupling. Firstly, we remove all nodes which are system components and edges connected to these nodes (e.g., the white nodes in the figure). Then, we obtain several disconnected subgraphs of the original RBG. Secondly, for each disconnected subgraph, we add back the removed system component nodes which have links with nodes in this subgraph. Then, we re-link the added nodes to nodes in the subgraph. Lastly, we will obtain several individual graphs (e.g., the two graphs in the upper circle and the lower dotted circle showed in the figure) which contain logic structure and runtime behavior belonging to these separated graphs. By using graph decoupling, we can easily separate malicious logic and runtime behaviors from original mixed RBG.

**(2) Malware Signature Generation:** Because malicious runtime behaviors are captured at runtime, some behaviors can only be triggered by certain events. Moreover, automatic app-behavior triggering is still an ongoing research problem, and existing studies [34], [35] cannot effectively trigger all malicious behaviors. To make the detection more accurate, malware analyzer should manually trigger the malicious events at runtime. Therefore, before matching an uploaded suspicious signature, malware analyzer needs to launch the captured malware samples in MONET and triggers the malicious behavior manually. MONET will generate RBG and SSS for this malware. For the RBG, MONET will then perform the graph decoupling process to obtain a set of individual RBGs. Malware analyzer then determines which RBG contains malicious behaviors. These malicious RBGs will be stored as malware signature in the signature database. In Section IV, we will elaborate the implementation of our signature database.

**(3) Signature Matching:** Signature matching is to match the uploaded suspicious runtime behavior signature (including SSS and RBG) with existing malware signatures in the database to determine whether an app is malware or not. The signature matching process consists of SSS matching and RBG matching. For SSS, suspicious system calls can be the indicator of a malware. For instance, one suspicious SSS contains a connection to a well-known remote command and control server, or it has an execution of a root exploit binary. For RBG matching, it involves two steps. In the first step, we use the graph decoupling algorithm to separate the suspicious RBG into a set of decoupled RBG ($\mathcal{D}$). For the second step, the backend detection server will execute a graph similarity algorithm to compare graph in the decoupled RBG set ($\mathcal{D}$) with graphs in the malware RBG set ($\mathcal{M}$). We say that there is a match if there exists a $d \in \mathcal{D}$ and an $m \in \mathcal{M}$ such that the similarity between $d$ and $m$ is smaller than a threshold ($\mathcal{T}$). In the MONET backend detection server, we use the graph edit distance algorithm to measure the similarity between two RBGs. The similarity of two runtime behavior graph $G_1$ and $G_2$ is: $sim(G_1, G_2) = 1 - \frac{\min(|V_i|+|V_d|+|E_i|+|E_d|)}{|V|+|V'|+|E|+|E'|}$, where $|V_i|$ and $|V_d|$ are the number of vertex-edit operations of vertex insertion and vertex deletion from $G_1$ to $G_2$. $|E_i|$ and $|E_d|$ are the number of edge-edit operations of vertex insertion and vertex deletion from $G_1$ to $G_2$. We calculate the minimum operation to transform $G_1$ to $G_2$. Then, $|V|+|V'|+|E|+|E'|$ quantifies the maximum operations from $G_1$ and $G_2$. Therefore, a high similarity score of two RBGs implies that it needs small number of transformations from one to another. Figure 6 illustrates an example of graph edit distance between two RBGs: $G_1$ and $G_2$. Both of them have six nodes and six edges. They have the same graph structure except that one edge in $G_2$ points to a different node (i.e., dotted link in the figure). The number of edge-edit operations from $G_1$ to $G_2$ is 2 because we have to delete one edge and insert a new edge. Therefore, the similarity score between $G_1$ and $G_2$ is $1 - (1 + 1 + 0 + 0)/(6 + 6 + 6 + 6) = 0.92$. In other words, these two graphs $G_1$ and $G_2$ is highly similar.

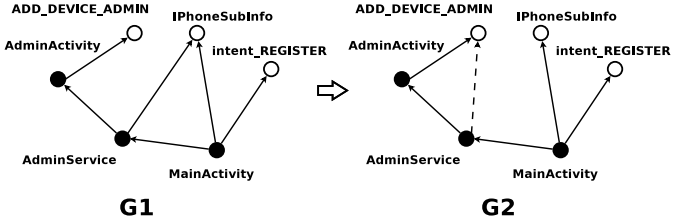

Fig. 6. Example of graph edit distance.

## IV. Implementation of Monet

In this section, we present the implementation of MONET. The system consists of two parts: a *client app* (which can be installed in any Android device) to capture the behavior and generate signatures, and a *backend detection server* to determine whether a suspicious app is a malware variant.

### A. Client App

The MONET client app can generate SBG for newly installed apps. At runtime, the MONET client app monitors intrusive transactions and system calls. Once a suspicious behavior is detected, the MONET client uses the collected runtime information to generate the RBG and the SSS for the executed app, and then sends them as the monitored behavior of that app to the backend detection server for malware detection. In our implementation, the client app consists of three main components, (1) *SBG generator*, (2) *runtime information collector* and (3) *RBG and SSS generator*.

**(1) SBG generator:** The MONET client app monitors the app installation events (i.e., PACKAGE_INSTALL and PACKAGE_ADDED action). Once an app is installed, SBG generator will use the smali/baksmali library [30] as a disassembler to disassemble newly installed apps. The output is a set of disassembled codes. In addition, the SBG generator will also translate the compiled binary AndroidManifest.xml file into a plain text file. As we discussed in Section III, to generate an SBG, the SBG generator will first generate a control flow graph (CFG) for each component class. Secondly, it will extract component information from the AndroidManifest.xml.

With the CFG and component information, it uses a data flow analysis technique and reaching definition algorithm to generate a static behavior graph based on compiler theory.

The reaching definition algorithm we used is based on the compiler theory, and the algorithm is depicted in Algorithm 1. Input to the algorithm is a CFG of an app component class generated from the disassembled code. In this algorithm, $GEN[B]$ is the definitions within the code block $B$, and $KILL[B]$ is the definitions which are redefined (i.e., assigned with other values) in block $B$. After calculating the reaching definition, we obtain two sets of definitions: $IN[B]$ and $OUT[B]$. $IN[B]$ contains definitions which reach $B$'s entry, and $OUT[B]$ contains definitions which reach $B$'s exit. For example, if we want to find the definition of variable `i` in the `startActivity(i)` block ($b$), using the reaching definition algorithm, we can obtain definitions that reach block B from $IN[b]$ list. If there is a definition of `i` in the list, we can find which statement defines the `i` variable. Lastly, we can also determine the value of `i` in that statement. In summary, this algorithm statically finds binder calls (links) between app components (nodes) to generate an SBG. The complexity of reaching definition algorithm is $O(n^2)$, where $n$ is the number of blocks in a CFG. For all the apps and malware we tested, the value of $n$ is between 1 and 20.

---
**Algorithm 1** Reaching definition algorithm
---
**Input:** Control flow graph: $CFG = (N, E, ENTRY, EXIT)$
**Output:** $IN[B]$ and $OUT[B]$ sets
  $OUT[ENTRY] \leftarrow \emptyset$
  **for all** basic block $B$ other than $ENTRY$ **do**
    $OUT[B] \leftarrow \emptyset$
  **end for**
  **while** changes to any $OUT$ occur **do**
    **for all** block $B$ other than $ENTRY$ **do**
      $IN[B] = \bigcup(OUT[p])$     ▷ for all predecessors $p$ of $B$
      $OUT[B] = GEN[B] \bigcup (IN[B] - KILL[B])$
    **end for**
  **end while**

---

**(2) Runtime Information Collector:** The runtime information collector runs in the background of an Android device and it collects all binder transactions to generate an RBG and specific system calls to generate an SSS. We implement the runtime information collector using two interception techniques on binder calls and system calls respectively. Figure 7 illustrates our implementation. It contains two functional parts: the *binder call interception* and the *system call interception*.

• **Binder Call Interception.** MONET needs to collect the binder calls information including the binder transaction code, the transaction descriptors and various additional attributes. The MONET client app uses the hooking technique on binder calls. In essence, the client app injects libraries into apps and system services to hook binder transactions. The first hooking place is on the JNI interface for intercepting upper binder related APIs between the Java layer and the native layer. Using this method, we can intercept *all binder calls* initiated by this app from the Java layer. The second hooking place is on the Service Manager. Because all binder requests need to first go through the Service Manager, the MONET client app will also intercept calls to the Service Manager to avoid any malware

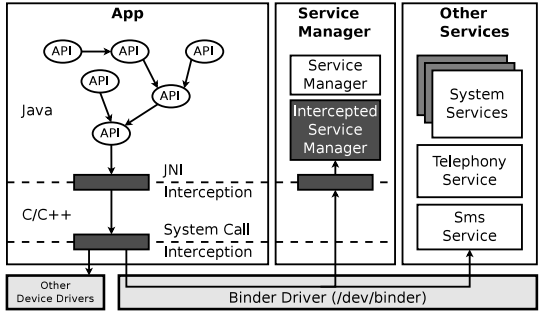

Fig. 7. Implementation of the MONET runtime information collector.

TABLE I
BINDER CALL INFORMATION AT RUNTIME.

| Caller Component | Target Component | Code | Code Action |
|---|---|---|---|
| *.MainActivity | PackageManager | 2 | getPackageInfo |
| *.WorkService | ConnectivityManager | 4 | getActiveNetworkInfo |
| *.WorkService | PhoneSubInfo | 4 | getDeviceId |
| *.AdminService | DevicePolicyManager | 41 | isAdminActive |
| ... | ... | ... | ... |

* Package name: com.google.elements

using native code to initiate malicious binder calls. Figure 7 depicts the technical details of our binder call interception. For example, if a malware uses the `sendTextMessage()` API to send a premium message, this API call will go through several lower layer APIs in the Android SDK. At the end, this method call will be handled by a binder object. This binder object will call the `transact()` JNI method to invoke the native function. MONET will capture this binder transaction and record this binder call. In addition, the MONET client app can also obtain the runtime calling stack trace of this JNI method to find out which component is initiating the binder call. Because this binder call is an intrusive transaction, we will then be able to notify users about the intrusive events. Note that MONET will also generate an RBG using the current collected runtime information and send it to the detection server for malware detection. Table I depicts some binder call records of the `o5android` malware. The record includes caller component names, target component names, binder call codes and corresponding actions of the codes. For example, the `com.google.elements.WorkService` component will request device ID from the PhoneInfoSub component at runtime. These binder records will be used to complete the SBG to generate RBG for detection.

• **System Call Interception.** To intercept system calls, we implement a loadable kernel module (LKM) for the Linux kernel. The kernel module will first search the address of the `sys_call_table` structure. The `sys_call_table` structure stores the pointers of system call implementations. In the MONET client app, we get the `sys_call_table` address from the `vector_swi` handler [36]. Using this method, we can determine the address of the `sys_call_table` for different build versions of the Linux kernel. Then, to intercept system calls, we change the system call addresses in the `sys_call_table` to addresses of our own functions.

Inside our methods, the MONET client app will write the calling information including caller process ID and system call parameters into a device driver (`/dev/monet`) to pass the information to the user layer app. At the end of the function, MONET will call back the original functions to continue the original logic of the app.

In our current implementation, we intercept two system calls: `socket()` (i.e., `sys_call_table[__NR_socket]`) and `execve()` (i.e., `sys_call_table[__NR_execve]`). By replacing the system call entries in the system call table, we redirect these two system calls to our interception first and then return back to their original system calls. For `execve()`, the kernel adds a wrapper to adjust the parameter `r3` before performing the actual `execve` task. The wrapper points `r3` to a stack location calculated from the stack pointer `sp`. Therefore, we should guarantee that the stack pointer `sp` is not corrupted during our interception.

Intercepting these two system calls can expose most of the malicious behavior in apps. Firstly, malware could use the network to communicate with their remote command and control servers. Therefore, to intercept this kind of behavior, we should intercept `socket()` system call in the kernel layer so that MONET will get the network connection information either from the Java APIs or from native codes. Secondly, many malware (e.g., `DroidKungfu`) attempt to execute a root exploit when launching the malware. Therefore, `execve()` is another dangerous behavior which we need to keep track.

We like to point out that the interception technique for binder calls is easy to deploy on Android devices. The deployment needs root privilege to inject libraries into apps and system services. There are several tools which provide root privilege management for apps. Moreover, they will also prevent malware abusing root privilege to keep the device secure. For the interception on system calls, because the kernel for the current Android system is stable and will not have many modifications, and loadable kernel module is compatible for the current systems and easy to deploy. Furthermore, using the above mentioned hooking technique, MONET can be deployed on a wide variety of Android-based mobile devices.

**(3) RBG and SSS Generator:** With the collected binder call and system call information, MONET builds an RBG and an SSS. RBG is based on the SBG which was generated at the installation time of a new app. From the runtime information collector, we can gain a vector of binder calls sequence at runtime with the caller class names, binder call descriptors and binder call content. With this information, we can complete the SBG to generate an RBG. For suspicious system calls, MONET reads the calling information from the kernel space via the device driver, and puts the system calls which belong to current app process to SSS.

### B. Backend Detection Server

The backend detection server is responsible for storing malware signatures in the database, and to perform malware detection using our signature matching algorithm. Because an SSS is for detecting network address and binary root exploit in the blacklist, the SSS matching is based on a traditional hashing matching implementation. Note that usually, we only need to use the RBG for the logic structure and runtime information for detection. The matching algorithm of RBG needs to perform graph similarity computation, but graph comparison is computationally expensive. Therefore, based on the properties of the runtime behavior graph, we use a B+ tree to index malware signature to optimize the detection process. In the current implementation, we use the number of app components as a key to the B+ tree, and this information is easy to derive from RBGs. To insert a record in the B+ tree, it only requires $O(\log_b n)$ operations, and performing a range query with $k$ elements requires $O(\log_b n+k)$ operations, where $n$ is the number of nodes in the B+ tree and $b$ is the maximum number of children nodes for the internal node. Lastly, by using the B+ tree, we only need to compare RBGs within a range. For example, if we need to detect an RBG with $n$ nodes, we only need to query and compare malware RBGs in our database within $n-\alpha$ and $n+\alpha$ nodes, where $\alpha$ is a constant integer we set in MONET. In our experiments, we set $\alpha = 5$. If the number of nodes for malware RBGs in the database is not in $[n-\alpha, n+\alpha]$, with high probability, the similarity scores between the uploaded RBG and RBGs in the database will be low. Using this method, it will reduce the comparison computation for malware detection.

Overall, the workflow of detection can be described as follows: (1) Monet detects suspicious transaction calls by monitoring IPC; (2) A warning dialog pops out to users and at the same time the signature is sent to server for evaluation; (3) Because these two operations are asynchronized processes, users can wait for detection results then decide whether to block the malicious events. Considering some detection may occur without network connection, we pre-loaded widely detected malware signatures for offline detection.

## V. EVALUATION

In this section, we first present our experimental setup and dataset. Then, we present the evaluation results on the accuracy and effectiveness of MONET to detect malware variants and defend against malware transformation. We also present the battery consumption of the MONET's client module.

### A. Experimental Setup & Dataset

In our experiment, we use an LG Nexus 5 mobile phone to test our client app. Our test phone runs the Google official Android firmware, or KitKat 4.4.4 with the build number KTU84P and kernel version 3.4.0. Our backend detection server is a Dual-core $3.10\,\text{GHz}$ PC and $8\,\text{GB}$ memory.

We collected 3,723 malware samples from the Android Malware Genome Project [20], DroidAnalytics [37] samples and contagio minidump forums [38]. In addition, we also downloaded the top 500 apps from the Google Play market (i.e., the ranking is based on the download number ranking list). Note that we need these legitimate apps to evaluate MONET's capability on true negative, as well as to explore the number of nodes within an RBG.

To analyze the characteristics of these apps, we execute these apps for one minute and generate their corresponding

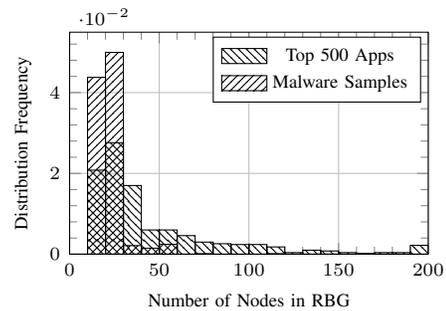

Fig. 8. Distribution of the number of nodes in RBG for top apps and malware samples.

RBGs. Figure 8 depicts the distribution of the number of nodes in an RBG for malware or for benign apps. From the figure, we see that most of the apps contain less than 50 nodes in their RBGs. In Section III, we discussed that many graph similarity algorithms require high computation. Because the number of nodes in RBG is small, the computation of graph comparison is therefore acceptable. We will present the performance evaluation of the backend detection server in later experiment results.

### B. Evaluation on Detection Capability

MONET uses the runtime behavior signature for malware detection. It can detect exiting malware samples and their variants, as well as malware which uses transformation techniques. Let us present our results.

**Experiment 1 (Accuracy and Effectiveness on Detecting Malware Variants):** `DroidKungfu` malware is a popular repackaged malware. It injects malicious classes into benign apps including tools and games. There are four variants (`DKF1`, `DKF2`, `DKF3` and `DKF4`) of `DroidKungfu` malware. The original malware (`DKF1`) listens to the battery change and boot complete actions. If these actions are triggered, `DKF1` performs several behaviors including reading/writing data in the XML file, starting another service, installing a new app, or gaining root privilege, etc. For the following evolved malware variants, `DKF2` uses native code to execute root exploit. `DKF3` uses string obfuscation and AES encryption methods to hide malicious string signature. `DKF4` uses the same package name as the hosted benign app to hide its static signatures.

We performed experiments to see the effectiveness of MONET in using one malware signature (e.g., `DKF1`) to detect other malware variants within the same malware family (e.g., `DKF2` to `DKF4`). Table II shows the detection results for each variant of the `DroidKungfu` malware family. We use 30 DFK1, 30 `DKF2`, 295 `DKF3` and 90 `DKF4` samples for detection. We measure the true positive (TP), false positive (FP), true negative (TN), false negative (FN) as well as the accuracy ($ACC = (TP+TN)/(TP+TN+FP+FN)$) for each `DroidKungfu` variant using SSS, or RBG only, or their combination as signature respectively. We set the threshold $\mathcal{T}$ to be $0.8$ for our detection server. For example, we first use one sample of `DroidKungfu1` to generate a runtime behavior signature. Then, we install *all other samples* and 500 benign apps on our test phone with MONET, and run the apps for one minute. To simulate user interactions, we use `monkey` [39] to generate 500 pseudorandom system/user events such as clicks, touches and gestures, etc. More sophisticated triggering methods or real users' interactions will help our system to capture runtime behavior thoroughly.

From our experiments, we found that 29 out of 30 are detected as `DKF1` malware, and so our true positive rate is $29/30$. There is one `DKF1` sample which is not detected as malware, so our false negative rate is $1/30$. We manually review the disassembled code of this malware sample. We found that hackers declare the malicious component name in the manifest file, but this malware *does not* contain any malicious logic. Because current anti-virus engines may depend on this unique static component name for detection. MONET is based on runtime behaviors, so this app will not be detected as malware. All 500 benign apps are detected correctly and so our true negative rate is 1, and none of the benign apps are reported as malware, so our false positive rate is 0. We also found that most malicious logic will be initiated at the startup time of malware samples. Therefore, one-minute running time is enough for performing this effectiveness evaluation. However, longer monitoring frames can help the system to comprehensively complete the runtime behaviors for detection.

Let us illustrate the effectiveness of MONET using the RGB and SSS for detection. From our experiment, we see that when using the runtime malware signature including RBG and SSS for detection, the average accuracy of detecting four `DroidKungfu` variants is around $99\%$. Secondly, if we only use RBG for detection, the accuracy is $98.5\%$, which drops a little but it is still very effective in malware variant detection. The reason is that some malicious binder calls and system calls are not triggered in the automatic triggering process. The average detection time on our test detection server is about $0.2$ seconds. The data transformation time through Wi-Fi network is about two seconds. In summary, the total detection time for each malware sample is less than three seconds under a stable network status.

Besides detecting existing malware within one variant, MONET is also effective to detect evolved malware variant. To illustrate this capability, we use a runtime behavior signature from one variant of the `DroidKungfu` family to detect other variants. Figure 9 illustrates the accuracy of our detection using different signatures. For example, we first use `DroidKungfu1` (`DKF1`) signature to detection other variant samples (`DKF2`, `DKF3`, `DKF4`). The accuracy for the next generation variant (`DKF2`) is still high. Because some samples of `DKF3` and `DKF4` variants change behavior in interacting with the command and control server, the detection accuracy drops a little. In summary, the detection accuracy of two consecutive variants is above $90\%$.

**Experiment 2 (Defending Against Malware Transformation):** Transformation attacks use static obfuscation tools to hide malicious logic. Traditional feature-based anti-virus engines rely heavily on specific patterns of malware for detection. But string obfuscation and encryption can change the pattern and bypass these transitional anti-virus engines. Moreover, obfuscation also makes the logic complicated such

TABLE II
DETECTION RESULTS FOR DROIDKUNGFU MALWARE FAMILY WITH 500 BENIGN APPS FROM GOOGLE PLAY.

| Malware Variants | # of Samples | SSS* | TPR | FNR | TNR | FPR | ACC |
|---|---|---|---|---|---|---|---|
| DKF1 | 30 | ○ | 0.10 | 0.90 | 1.00 | 0.00 | 94.9% |
|  |  | ◐ | 0.97 | 0.03 | 1.00 | 0.00 | 99.8% |
|  |  | ● | 0.97 | 0.03 | 1.00 | 0.00 | 99.8% |
| DKF2 | 30 | ○ | 0.33 | 0.67 | 1.00 | 0.00 | 96.2% |
|  |  | ◐ | 1.00 | 0.00 | 1.00 | 0.00 | 100.0% |
|  |  | ● | 1.00 | 0.00 | 1.00 | 0.00 | 100.0% |
| DKF3 | 295 | ○ | 0.11 | 0.89 | 1.00 | 0.00 | 69.9% |
|  |  | ◐ | 0.99 | 0.01 | 1.00 | 0.00 | 99.6% |
|  |  | ● | 0.92 | 0.08 | 1.00 | 0.00 | 96.9% |
| DKF4 | 90 | ○ | 0.18 | 0.82 | 1.00 | 0.00 | 87.6% |
|  |  | ◐ | 0.98 | 0.02 | 1.00 | 0.00 | 99.7% |
|  |  | ● | 0.89 | 0.11 | 1.00 | 0.00 | 98.3% |
| **Total** | 445 | ○ | 0.14 | 0.86 | 1.00 | 0.00 | 87.2% |
|  |  | ◐ | 0.99 | 0.01 | 1.00 | 0.00 | 99.7% |
|  |  | ● | 0.92 | 0.08 | 1.00 | 0.00 | 98.5% |

* Runtime behavior signature usage: ○: SSS, ◐: RBG only
●: SSS and RBG together.

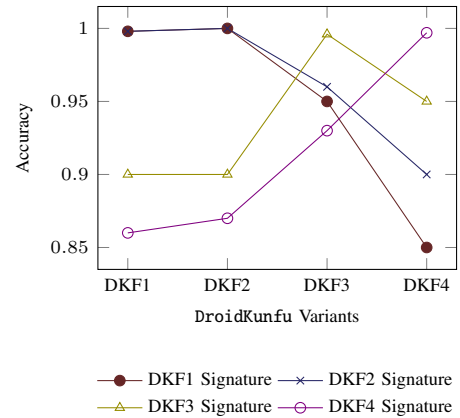

Fig. 9. Detecting `DroidKungfu` Malware Variants.

that malware researchers cannot easily analyze the malicious logic. Instead of relying on string patterns, MONET uses malicious behaviors for detection because malicious behaviors are difficult to transform. In this experiment, we use a self-made malware (`o5android`). This malware will request for device administrator, or send text messages, or gain device id, etc. Moreover, hackers generated a set of malware which have a random configuration file so the MD5 values are different. We also use two transformation tools (ADAM [22] and DroidChameleon [23]) to generate 45 obfuscated apps from three original malware. In addition, we also implement reflection and dynamic loading techniques to complement existing methods. We use twelve types of transformation techniques in the experiment. Table III shows the descriptions of these twelve transformation techniques. We install these 45 transformed malware on the device with the MONET client module. 40 out of 45 are detected as `o5android` malware by our system. Because some techniques used by the transformation tools may corrupt the logic of malware, five of them crash after transformation. So we cannot consider them in the experiment. We also conduct an experiment on a real world malware family called `FakeAV`. This malware family utilizes a simple transformation method to generate large amount of samples. We successfully detect all collected nine malware samples with different hash values (e.g., SHA1).

TABLE III
DESCRIPTIONS OF TRANSFORMATION TECHNIQUES.

| Transformation Techniques | # of Samples | # of Detection |
|---|---|---|
| 1. renaming classes | 6 | 5 |
| 2. reversing bytecode order | 3 | 3 |
| 3. string encryption | 6 | 5 |
| 4. arrays encryption | 3 | 3 |
| 5. removing debug information | 3 | 3 |
| 6. reordering instructions | 3 | 3 |
| 7. inserting non-trivial junk instructions | 6 | 5 |
| 8. inserting NOP instructions | 3 | 3 |
| 9. renaming method | 6 | 5 |
| 10. renaming fields | 6 | 5 |
| 11. reflection | 3 | 3 |
| 12. dynamic loading | 2 | 2 |
| **Total** | 50 | 45 |

TABLE IV
BENCHMARK RESULTS.

| Test | Baseline | MONET | Overhead |
|---|---|---|---|
| **CPU** | 21 043 | 20 015 | 4.8 % |
| **Memory** | 14 201 | 13 019 | 8.3 % |
| **I/O** | 7334 | 6782 | 7.5 % |
| **2D** | 325 | 311 | 4.0 % |
| **3D** | 2320 | 2302 | 0.8 % |
| **Composite** | 8802 | 8142 | 7.4 % |

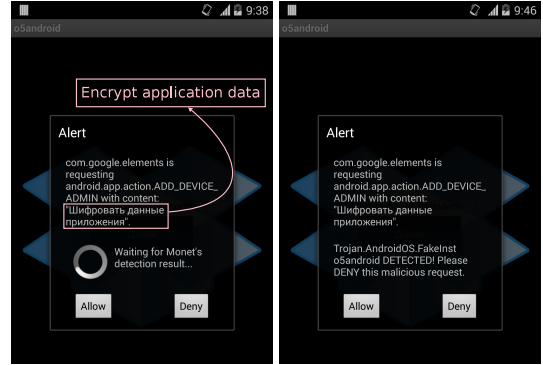

Fig. 10. Screenshots of MONET.

**Experiment 3 (Performance and Battery Overhead):** We use Quadrant Standard Edition v2.1.1 [40] to measure the general purpose benchmark for CPU, memory, I/O, 2D and 3D graphics. Table IV shows the benchmark results. Because MONET will intercept binder calls and system calls, we have round 8 % overhead in memory and I/O benchmarks. We also measure the battery overhead introduced by MONET. We first check the battery overhead in the standby mode. We use a fully-charged test phone in standby mode for 24 hours. The device with MONET installed only has 3.2 % battery overhead as compared with device without MONET. Then, we use the phone for one hour with heavy usage including 20 minutes game playing, 20 minutes network surfing and 20 minutes telephone call. We monitor the battery capacity by reading the `/sys/class/power_supply/battery/capacity` file. The battery of MONET for a heavy user is about 5.5 %. In summary, MONET has a low impact on the battery resource.

**Experiment 4 (Capability to Alert Users):** Figure 10 demonstrates two screenshots of MONET. When users launch the `o5android` malware, MONET detects a malicious behavior, which is requesting users to add itself as a device administration. From the left screenshot, MONET shows a popup dialog to indicate the app is starting the device manager for `ADD_DEVICE_ADMIN` action. The content of this intent is a message in Russian which means "encrypt application data". `o5android` is using this message to deceive users to accept this `ADD_DEVICE_ADMIN` request. At the same time of this alert, MONET will send runtime behavior signature to the backend detection server. In the right screenshot, the alert dialog shows the detection result, and users can click "Deny" button to avoid executing malicious behavior.

## VI. RELATED WORK

With the emergence of malware on the Android ecosystem, researchers have proposed a number of systems to detect Android malware based on static resources such as permission information, disassembled codes and other resources. Zhou et al. [41] and Asokan [42] systematically analyze the evolution of Android malware. DroidMOSS [20], Juxtapp [43], DNADroid [44], AnDarwin [45], MassVet [46], ViewDroid [47], Dendroid [48], ResDroid [49], and DroidEagle [50] aim at detecting repackaged and clone malware. DroidRanger [21] uses permission-based footprinting and heuristic schemes to detect existing malware. RiskRanker [32] can automatically uncover malicious behaviors of zero-day malware. DREBIN [51], DroidSIFT [52] and ICCDetector [53] use machine learning algorithm to detect malware. There are a number of works [19], [54]–[57] which use static dataflow analysis to identify malicious logic in Android apps and classify existing malware. To prevent malware exploiting capability leaks and content leaks vulnerabilities, systems [11], [20] aim at detecting such loopholes in apps. All these systems are based on static features of malware. However, current malware use advanced obfuscation methods to bypass disassembled tools or hide the malicious logic in native code. Moreover, learning-based malware algorithm is not computational efficiency and their effectiveness strongly depends on the feature selection. In contrast, our system uses *both* static features and dynamic runtime information to describe malicious behavior, and MONET is effective in defense against logic transformation.

To analyze sophisticated malware, researchers propose a number of dynamic analysis systems. TaintDroid [58], TaintART [59], DroidScope [60], VetDroid [61], CopperDroid [62] and DroidBox [63] detect malicious behavior using dynamic analysis. Marvin [64] combines static and dynamic analysis for classifying malicious apps. In addition, some systems [65] are proposed to track information flow to prevent privacy leakage. However, these systems are designed for malware analysts. It is difficult for regular mobile device users to install them on their device to detect and prevent malware. Therefore, several systems [27]–[29], [66]–[70] are proposed to prevent intrusion on devices for regular users. However, these systems can only warn users about the suspicious behaviors at runtime, and users cannot easily determine whether a suspicious behavior is from a malware or not. Our system is designed for regular mobile users. If an intrusion from a suspicious app is detected, MONET can effectively determine the malware from its runtime behavior and alert the user.

There are a number of malware detection systems based on dynamic behavior or runtime information for mobile devices. Bose et al. [71] propose a behavior-based detection system for Symbian OS, which is an outdated mobile system. At that time, malware in mobile devices were rare and simple. pBMDS [72] and DroidScribe [73] uses machine learning methods to classify the behaviors of apps. However, the model only works on keyboard inputs, while most interactions with devices are on the touchscreen nowadays. Crowdroid [74] and MADAM [75] utilize system call sequences as malware behavior for detection. System calls contain less semantic information and cannot accurately represent a malicious behavior. MONET captures binder transactions and system calls, for they contain more semantic information which can accurately describe the runtime behavior.

## VII. CONCLUSION

In this paper, we present the design and implementation of MONET to detect malware variants and to defend against transformation attack. MONET will generate a runtime behavior signature which consists of RBG and SSS to accurately represent the runtime behavior of a malware. Our system includes a backend detection server and a client app which is easy to deploy on mobile devices. Our experiments show that MONET can accurately detect malware variants and defend against transformation attacks with only a minimal performance and battery overhead. Note that recently, Google released Android 5.0 Lollipop which will replace the Dalvik virtual machine with ART. ART runtime abandons the virtual machine mechanism, but uses the ahead-of-time compilation. Therefore, our current implementation using the binder interception may not be directly applicable to the ART runtime. However, because the application package structure and binder mechanism remain unchanged, so one can easily extend MONET on the ART runtime. This is our future work.

TECHNICAL REPORT OF MONET: A USER-ORIENTED BEHAVIOR-BASED MALWARE VARIANTS DETECTION SYSTEM FOR ANDROID 13